\documentclass[12pt]{article}
\usepackage{axodraw}

\catcode`@=12
\begin{document}
\thispagestyle{empty}
\begin{flushright} 
UCRHEP-T355\\ 
June 2003\
\end{flushright}
\vspace{0.5in}
\begin{center}
{\LARGE	\bf Different $G_F$ and $\sin^2 \theta_W$ for different processes\\}
\vspace{1.5in}
{\bf Ernest Ma\\}
\vspace{0.2in}
{\sl Physics Department, University of California, Riverside, 
California 92521\\}
\vspace{2.0in}
\end{center}
\begin{abstract}\
A natural theoretical framework is presented which allows for small departures 
from quark-lepton universality such that different $G_F$ and $\sin^2 \theta_W$ 
values are applicable for different processes.  In particular, the inequality 
$(G_F)^{NC}_{lq} < (G_F)^{CC}_{lq} < (G_F)^{CC}_{ll} < (G_F)^{NC}_{ll}$ holds. 
New physics is predicted at the TeV scale.
\end{abstract}

---------

Preprint version of talk at NUINT 02

\newpage
\baselineskip 24pt

\section{Introduction}

In the Standard Model, the low-energy effective weak interactions are of 
the form
\begin{equation}
\frac{4 G_F}{\sqrt 2} \left[ j^{(+)} j^{(-)} + \left( 
j^{(3)} - \sin^2 \theta_W j^{(em)} \right)^2 \right],
\end{equation}
where
\begin{equation}
\frac{4 G_F}{\sqrt 2} = \frac{g^2}{2 M_W^2} = \frac{g^2 + g'^2}{2 M_Z^2} = 
\frac{1}{v^2}.
\end{equation}
Note that $G_F$ is independent of $g$ and $g'$.

As a result of Eq.~(1), there are 3 predictions:
\begin{eqnarray}
&(A)& G_F^q = G_F^l, ~~ \sin^2 \theta_W^q = \sin^2 \theta_W^l; \\ 
&(B)& G_F^e = G_F^\mu = G_F^\tau; \\
&(C)& G_F^{CC} = G_F^{NC}.
\end{eqnarray}
Possible experimental deviations of (A) and (C) have now been observed at the 
3$\sigma$ level.  Whereas it is too early to tell for sure that these are real 
effects, it is clearly desirable to have a theoretical framework where 
departures from quark-lepton universality are naturally expected and which 
reduces to the Standard Model in the appropriate limit.

\section{Three Experimental Discrepancies}

\noindent (1) A recent measurement \cite{neutron} of the neutron $\beta-$decay 
asymmetry has determined that
\begin{equation}
|V_{ud}| = 0.9713(13),
\end{equation}
which, together with \cite{pdg} $|V_{us}| = 0.2196(23)$ and $|V_{ub}| = 
0.0036(9)$, implies the apparent nonunitarity of the quark mixing matrix, i.e.
\begin{equation}
|V_{ud}|^2 + |V_{us}|^2 + |V_{ub}|^2 = 0.9917(28).
\end{equation}
However, if $(G_F)^{CC}_{lq} < (G_F)^{CC}_{ll}$, as we will show, then the 
above is actually \underline {expected}.

\noindent (2) The NuTeV experiment \cite{nutev} which measures $\nu_\mu$ and 
$\bar \nu_\mu$ scattering on nucleons reported a value of
\begin{equation}
\sin^2 \theta_W = 0.2277 \pm 0.0013 \pm 0.0009,
\end{equation}
where the Standard-Model expectation is $0.2227 \pm 0.00037$, assuming 
$(G_F)^{NC}_{lq}/(G_F)^{CC}_{lq} = 1$.  In our model, this ratio will be 
smaller than one, which would explain the data if it is $0.9942 \pm 0.0013 
\pm 0.0016$ and $\sin^2 \theta_W$ does not change.  However, we do expect the 
latter to change, but since its precise determination comes from $Z$ decay, 
we need to consider also data at the $Z$ resonance.

\noindent (3) In precision measurements of $e^- e^+ \to Z \to q \bar q$ 
and $l \bar l$, there seem to be two different values of 
$\sin^2 \theta_{eff}$, i.e. \cite{ichep}
\begin{eqnarray}
(\sin^2 \theta_{eff})_{hadrons} &=& 0.23217(29), \\ 
(\sin^2 \theta_{eff})_{leptons} &=& 0.23113(21).
\end{eqnarray}
This may be an indication of a small deviation from quark-lepton 
universality.

In this talk I will show that (1) is naturally explained by a gauge model of 
quark-lepton nonuniversality \cite{lima02}, the prototype of which was 
proposed over 20 years ago \cite{lima81} for generation nonuniversality. 
As a result, effects indicated by (2) and (3) are also expected, but the 
observed deviations are too large.

\section{Gauge Model of Quark-Lepton Nonuniversality}

Consider the gauge group $SU(3)_C \times SU(2)_q \times SU(2)_l \times 
U(1)_q \times U(1)_l$ with couplings $g_s$ and $g_{1,2,3,4}$ respectively. 
The quarks and leptons transform as
\begin{eqnarray}
(u,d)_L &\sim& (3,2,1,1/6,0), \\ 
u_R &\sim& (3,1,1,2/3,0), \\ 
d_R &\sim& (3,1,1,-1/3,0), \\ 
(\nu,e)_L &\sim& (1,1,2,0,-1/2), \\ 
e_R &\sim& (1,1,1,0,-1).
\end{eqnarray}
The scalar sector consists of
\begin{eqnarray}
(\phi_1^+,\phi_1^0) &\sim& (1,2,1,1/2,0), \\ 
(\phi_2^+,\phi_2^0) &\sim& (1,1,2,0,1/2), \\ 
\chi^0 &\sim& (1,1,1,1/2,-1/2),
\end{eqnarray}
and a bidoublet
\begin{equation}
\eta = \frac{1}{\sqrt 2} \pmatrix {\eta^0 & -\eta^+ \cr  
\eta^- & \bar \eta^0} \sim (1,2,2,0,0),
\end{equation}
which is assumed to be self-dual, i.e. $\eta = \tau_2 
\eta^* \tau_2$. Note that $g_1$ may be different from $g_2$, and $g_3$ may 
be different from $g_4$, so there is no quark-lepton symmetry at this level. 
The remarkable fact is that the effective low-energy weak interactions of 
the quarks and leptons will turn out to be independent of $g_{1,2,3,4}$ 
and become all equal in a certain limit, as shown below.

Consider
\begin{equation}
\langle \phi_{1,2}^0 \rangle = v_{1,2}, ~~ \langle \chi^0 \rangle = w, ~~ 
\langle \eta^0 \rangle = u,
\end{equation}
then the $2 \times 2$ charged-gauge-boson mass-squared matrix is given by
\begin{equation}
{\cal M}^2_W = \frac{1}{2} \left[ \begin{array} {c@{\quad}c} g_1^2 (v_1^2 + 
u^2) & -g_1 g_2 u^2 \\ -g_1 g_2 u^2 & g_2^2 (v_2^2 + u^2) \end{array} \right].
\end{equation}
Thus the effective lepton-lepton charged-current weak-interaction strength, 
i.e. that of $\mu$ decay, is
\begin{eqnarray}
&& \frac{4(G_F)^{CC}_{ll}}{\sqrt 2} = \frac{g_2^2}{2} \left( 
{\cal M}_W^{-2} \right)_{22} \nonumber \\ 
&&  = \frac{u^2+v_1^2}{(v_1^2+v_2^2)u^2+v_1^2v_2^2},
\end{eqnarray}
whereas the analogous expression for nuclear $\beta$ decay is
\begin{eqnarray}
&& \frac{4(G_F)^{CC}_{lq}}{\sqrt 2} = \frac{g_1g_2}{2} \left( 
{\cal M}_W^{-2} \right)_{12} \nonumber \\ 
&& = \frac{u^2}{(v_1^2+v_2^2)u^2+v_1^2v_2^2}.
\end{eqnarray}
Note that both are independent of $g_1$ and $g_2$, and their ratio is not 
one, but rather
\begin{equation}
\frac{(G_F)^{CC}_{lq}}{(G_F)^{CC}_{ll}} = \frac{u^2}{u^2+v_1^2} \simeq 1 - 
\frac{v_1^2}{u^2}.
\end{equation}
The apparent nonunitarity of the quark mixing matrix, i.e. Eq.~(7), is then 
naturally explained with
\begin{equation}
\frac{v_1^2}{u^2} = 0.0042(14).
\end{equation}
As for the effective neutral-current interactions, we have
\begin{eqnarray}
&& \frac{4(G_F)^{NC}_{lq}}{\sqrt 2} = \frac{u^2 w^2}{(v_1^2+v_2^2)u^2w^2 + 
v_1^2 v_2^2(u^2+w^2)} \nonumber \\ && \simeq \frac{4(G_F)_\mu}{\sqrt 2} \left[ 
1 - \frac{v_1^2}{u^2} - \left( \frac{v_2^2}{v_1^2+v_2^2} \right) \frac{v_1^2}
{w^2} \right], \\
&& \frac{4(G_F)^{NC}_{ll}}{\sqrt 2} = \frac{u^2 w^2 + v_1^2(u^2+w^2)}
{(v_1^2+v_2^2)u^2w^2 + v_1^2 v_2^2(u^2+w^2)} \nonumber \\ && \simeq
\frac{4(G_F)_\mu}{\sqrt 2} \left[ 1 + \left( 
\frac{v_1^2}{v_1^2+v_2^2} \right) \frac{v_1^2}{w^2} \right].
\end{eqnarray}
This implies that the ratio
\begin{equation}
\frac{(G_F)^{NC}_{lq}}{(G_F)^{CC}_{lq}} \simeq 1- \left( \frac{v_2^2}{v_1^2
+v_2^2} \right) \frac{v_1^2}{w^2}
\end{equation}
is what NuTeV actually measures \cite{nutev}. The corresponding $\sin^2 
\theta_W$ expressions depend on the identification of the observed $Z$ boson 
as a linear combination of the 3 massive neutral gauge bosons of this model, 
which will be discussed in the next section. 

\section{Observables at the $Z$ Pole}

There are 4 electroweak gauge couplings in this model.  The electromagnetic 
coupling $e$ is given by
\begin{equation}
\frac{1}{e^2} = \frac{1}{g_1^2} + \frac{1}{g_2^2} + \frac{1}{g_3^2} + 
\frac{1}{g_4^2}.
\end{equation}
Defining $g_{ij}^{-2} \equiv g_i^{-2} + g_j^{-2}$, the photon $A$ and 3 
orthonormal $Z$ bosons are given in the basis $(W_q^0, W_l^0, B_q, B_l)$ by
\begin{eqnarray}
A &=& e \left( \frac{1}{g_1}, \frac{1}{g_2}, \frac{1}{g_3}, \frac{1}{g_4} 
\right), \\ 
Z_1 &=& e \left( \frac{g_{12}}{g_{34}g_1}, \frac{g_{12}}{g_{34}g_2}, 
\frac{-g_{34}}{g_{12}g_3}, \frac{-g_{34}}{g_{12}g_4} \right), \\ 
Z_2 &=& g_{12} \left( \frac{1}{g_2}, \frac{-1}{g_1}, 0, 0 \right), \\ 
Z_3 &=& g_{34} \left( 0, 0, \frac{1}{g_4}, \frac{-1}{g_3} \right).
\end{eqnarray}
The observed $Z$ boson is approximately $Z_1 - \epsilon_2 Z_2 - 
\epsilon_3 Z_3$, where
\begin{eqnarray}
\epsilon_2 &\simeq& \frac{g_{34} g_{12}^4}{e g_1^3 g_2^3} \left( \frac{g_1^2 
v_1^2 - g_2^2 v_2^2}{u^2} \right), \\ 
\epsilon_3 &\simeq& \frac{g_{12} g_{34}^4}{e g_3^3 g_4^3} \left( \frac{-g_3^3 
v_1^2 + g_4^2 v_2^2}{w^2} \right).
\end{eqnarray}
Deviations from the Standard Model must occur and quark-lepton universality 
in $Z$ decay is violated if $\epsilon_2 \neq 0$ or $\epsilon_3 \neq 0$.

We have obtained \cite{lima02} all the appropriate expressions for the 
expected deviations from the Standard Model in terms of 5 parameters:
\begin{equation}
\frac{v_1^2}{u^2}, ~\frac{v_1^2}{w^2}, ~r \equiv \frac{v_2^2}{v_1^2}, 
~y \equiv \frac{g_2^2}{g_1^2+g_2^2}, ~x \equiv \frac{g_4^2}{g_3^2+g_4^2},
\end{equation}
and performed a global fit to 22 observables.  The best-fit values are 
\begin{eqnarray}
&& \frac{v_1^2}{u^2} = 0.00489, ~~ \frac{v_1^2}{w^2} = 0.00238, \\ 
&& r = 10.2, ~~ y = 0.0955, ~~ x = 0.135.
\end{eqnarray}
Our results are summarized in Table I.

\begin{table*}[htb]
\caption{Fit Values of 22 Observables}
\begin{tabular}{cccccc}
\hline
Observable & Measurement & Standard Model & Pull & This Model & Pull \\ 
\hline
$\Gamma_l$ [MeV] & $83.985 \pm 0.086$ & 84.015 & $-0.3$ & 83.950 & $+0.4$ \\ 
$\Gamma_{inv}$ [MeV] & $499.0 \pm 1.5$ & 501.6 & $-1.7$ & 501.2 & $-1.5$ \\ 
$\Gamma_{had}$ [GeV] & $1.7444 \pm 0.0020$ & 1.7425 & $+1.0$ & 1.7444 & $-0.0$ 
\\ 
$A^{0,l}_{fb}$ & $0.01714 \pm 0.00095$ & 0.01649 & $+0.7$ & 0.01648 & $+0.7$ 
\\ 
$A_l(P_\tau)$ & $0.1465 \pm 0.0032$ & 0.1483 &$-0.6$ & 0.1482 & $-0.5$ \\ 
$R_b$ & $0.21644 \pm 0.00065$ & 0.21578 & $+1.0$ & 0.21582 & $+1.0$ \\ 
$R_c$ & $0.1718 \pm 0.0031$ & 0.1723 & $-0.2$ & 0.1722 & $-0.1$ \\ 
$A_{fb}^{0,b}$ & $0.0995 \pm 0.0017$ & 0.1040 & $-2.6$ & 0.1039 & $-2.6$ \\ 
$A_{fb}^{0,c}$ & $0.0713 \pm 0.0036$ & 0.0743 & $-0.8$ & 0.0740 & $-0.8$ \\ 
$A_b$ & $0.922 \pm 0.020$ & 0.935 & $-0.7$ & 0.934 & $-0.6$ \\ 
$A_c$ & $0.670 \pm 0.026$ & 0.668 & $+0.1$ & 0.665 & $+0.2$ \\ 
$A_l$(SLD) & $0.1513 \pm 0.0021$ & 0.1483 & $+1.4$ & 0.1482 & $+1.5$ \\ 
$\sin^2 \theta^{lept}_{eff}(Q_{fb})$ & $0.2324 \pm 0.0012$ & 0.2314 & $+0.8$ 
& 0.2322 & $+0.2$ \\
$m_W$ [GeV] & $80.449 \pm 0.034$ & 80.394 & $+1.6$ & 80.390 & $+1.7$ \\ 
$\Gamma_W$ [GeV] & $2.139 \pm 0.069$ & 2.093 & $+0.7$ & 2.093 & $+0.7$ \\ 
$g_V^{\nu e}$ & $-0.040 \pm 0.015$ & $-0.040$ & $-0.0$ & $-0.039$ & $-0.1$ \\ 
$g_A^{\nu e}$ & $-0.507 \pm 0.014$ & $-0.507$ & $-0.0$ & $-0.507$ & $-0.0$ \\ 
$(g_L^{eff})^2$ & $0.3001 \pm 0.0014$ & 0.3042 & $-2.9$ & 0.3032 & $-2.2$ \\ 
$(g_R^{eff})^2$ & $0.0308 \pm 0.0011$ & 0.0301 & $+0.6$ & 0.0299 & $+0.8$ \\ 
$Q_W$(Cs) & $-72.18 \pm 0.46$ & $-72.88$ & $+1.5$ & $-72.26$ & $+0.2$ \\ 
$Q_W$(Tl) & $-114.8 \pm 3.6$ & $-116.7$ & $+0.5$ & $-115.7$ & $+0.3$ \\ 
$\sum_{i=d,s,b} |V_{ui}|^2$ & $0.9917 \pm 0.0028$ & 1.0000 & $-3.0$ & 0.9902 
& $+0.5$ \\
\hline
\end{tabular}
\end{table*}

We see that we are able to explain the apparent nonunitarity \cite{neutron} 
of the quark mixing matrix and reduce the NuTeV discrepancy \cite{nutev} 
while maintaining excellent agreement with precision data at the $Z$ 
resonance, except for the $b \bar b$ forward-backward asymmetry measured 
at LEP, which is also not explained by the standard model.  In fact, the 
shift of $A_{fb}^{0,b}$ is given in our model by
\begin{eqnarray}
\Delta A_{fb}^{0,b} &=& \frac{3}{4} (A_e \Delta A_b + A_b \Delta A_e) 
\nonumber \\ 
&=& -0.07 \Delta \sin^2 \theta_q - 5.57 \Delta \sin^2 \theta_l.
\end{eqnarray}
Because of the dominant coefficient of the second term, it measures 
essentially the same quantity as $A_l$ and there is no realistic means of 
reconciling the discrepancy of $\sin^2 \theta_{eff}$ at the $Z$ resonance 
using $b \bar b$ versus using leptons in the final state.

\section{Other Effects}

The new polarized $e^-e^- \to e^-e^-$ experiment (E158) at SLAC 
(Stanford Linear Accelerator Center) is designed to measure the left-right 
asymmetry which is proportional to $G_F (1-4\sin^2 \theta_W)$ to an accuracy 
of about 10\%.  Using the standard-model prediction of 
$\sin^2 \theta_W = 0.238$, our expectation is that the above measurement will 
shift by only $-2.2$\% from its standard-model prediction.  The new polarized 
$ep$ elastic scattering experiment (Qweak) at TJNAF (Thomas 
Jefferson National Accelerator Facility) is designed to measure $Q_W$ of 
the proton to an accuracy of about 4\%.  We expect a shift of only $+3.0$\%. 
Using Eq.~(37), we see also that the scale of new physics, i.e. $u$ and $w$, 
is at the TeV scale.  Specifically, using the best-fit values of $r$, $y$, 
and $x$, we find $M_{W_2} \simeq M_{Z_2} \simeq 1.2$ TeV, and 
$M_{Z_3} \simeq 0.8$ TeV.

\section*{Acknowledgments}

This work was supported in part by the U.~S.~Department of Energy under Grant 
No.~DE-FG03-94ER40837.


\end{document}